\documentclass[pra,twocolumn,epsfig,graphics,showpacs,floatfix,mathbbm]{revtex4}
\usepackage{amsmath,times}
\usepackage{psfrag}
\begin{document}
\title{Quantum controlled phase gate and cluster states generation via two superconducting
quantum interference devices in a cavity}
\author{Zheng-Yuan Xue$^{1}$\footnote{E-mail: xuezhengyuan@yahoo.com.cn},
Gang Zhang$^{1,2}$, Ping Dong$^{1}$, You-Min Yi$^{1}$ and Zhuo-Liang
Cao$^{1}$\footnote{E-mail: zlcao@ahu.edu.cn}}

\affiliation{1 School of Physics \& Material Science, Anhui
University, Hefei, 230039, P. R. China\\
2 Department of Mathematics \& Physics, West Anhui University,
Lu'an, 237012, P. R. China}

\begin{abstract}
A scheme for implementing 2-qubit quantum controlled phase gate
(QCPG) is proposed with two superconducting quantum interference
devices (SQUIDs) in a cavity. The gate operations are realized
within the two lower flux states of the SQUIDs by using a quantized
cavity field and classical microwave pulses. Our scheme is achieved
without any type of measurement, does not use the cavity mode as the
data bus and only requires a very short resonant interaction of the
SQUID-cavity system. As an application of the QCPG operation, we
also propose a scheme for generating the cluster states of many
SQUIDs.
\end{abstract}


\maketitle

\section{Introduction}
Recently, much attention has been paid to the realization of quantum
computers \cite{qcomputer}, which could compete in certain tasks
that classical computer could never fulfill in acceptable times
\cite{time}. Despite the already rather advanced theoretical
concepts of quantum computing, the development of its physical
implementations is just at an early stage. Up to now, many physical
systems have been suggested as possible realizations of qubits and
quantum logic gates \cite{proposals}. In some mature systems quantum
manipulations of a few qubits have already been demonstrated
experimentally, such as cavity QED \cite{qed}, trapped ions
\cite{ion}, nuclear magnetic resonance (NMR) \cite{nmr} and photonic
systems \cite{photon}. Despite recently exciting experimental
progresses, the physical realization of a scalable quantum computer
remains a great challenge. The building blocks of quantum computers
are two-qubit logic gates. Few systems have demonstrated controlled
qubits and qubit coupling between pairs taken from more than four
qubits. It is difficult to couple different subsystems in a
controlled manner, while at the same time shielding the system from
the influence of its environment. One particularly attractive
possibility is to use matter qubits to serve as hardware because
they are static and potentially long lived, and an optical coupling
mechanism can creates suitable entanglement. Among the variety of
systems being explored for hardware implementations of quantum
computers, cavity QED system is favored because of its demonstrated
advantage when subjected to coherent manipulations \cite{qed}.
However, in the cavity-atom system the strong coupling limit is
difficult to meet and individual addressing of the particles is also
a problem-maker. In contrast, the strong coupling limit was realized
with superconducting charge \cite{charge} and flux qubits
\cite{flux} in a microcavity. SQUIDs, as well as other solid state
circuits, can be perfectly fixed, easily embedded in a cavity and
easy to scale up, thus the cavity-SQUID scheme may be more
preferable than cavity-atom system. In addition, by placing SQUIDs
into a superconducting cavity, the external environment induced
decoherence can be greatly suppressed because the cavity can serve
as magnetic shield for SQUIDs.

It is well known that single-qubit rotations and 2-qubit controlled
logic gates together can be served to realize any unitary operation
on $n$ qubits \cite{qcomputing}. Recently, people have presented
various methods for implementing quantum logic operation via SQUID
flux qubits. Zhou \emph{et al}. \cite{3} proposed a scheme to
implement a single-qubit operation with a three-level $\Lambda$-type
rf-SQUID qubit, which proven to be more favorable than the
conventional two-level qubit. Amin \emph{et al}. \cite{4} gave a
more general method to implement an arbitrary qubit rotation using
the three-level qubit. However, under their assumption of small
detuning, population in the upper level of the qubit is significant
during the interaction, which results in higher probability of
spontaneous decay, and hence gate errors. Yang \textit{et al}.
\cite{5} proposed an alternative method for implementing arbitrary
single-qubit operations with a three-level SQUID without energy
relaxation. The above works \cite{3,4,5} have proposed some schemes
for realizing single-qubit operation and two-qubit entanglement, but
the building blocks of quantum computers are two-qubit logic gates.
Very recently, Song \emph{et al}. \cite{6} realized the 2-qubit QCPG
operation by using a quantized cavity field and classical microwave
pulses via Raman transition. In their scheme three SQUID qubits were
used to perform the QCPG operation, one of the SQUID qubits is
utilized as a quantum data-bus, and the gate operation is achieved
between the remaining two SQIUID qubits. Yang \emph{et al}. \cite{7}
proposed their scheme for realizing the n-qubit ($n\geq2$) QCPG
operation by using SQUIDs coupled to a superconducting resonator,
where data-bus qubit of the gate operation is not needed.
Enlightened by their progress, we report alternative ways of
demonstrating 2-qubit QCPG operation via SQUIDs in a quantized
superconducting cavity. Our scheme is achieved without any type of
measurement, does not use the cavity mode as the data bus and only
requires a very short resonant interaction.

In most of the current explorations, quantum logic gates are
implemented with sequences of controlled interactions between
selected particles. Recently, Raussendorf and Briegel \cite{oneway}
proposed a different model of a scalable quantum computer, namely
the one way quantum computer, which constructs quantum logic gates
by single-qubit measurements on cluster states. Thus many quantum
computation scheme based on the cluster states have been proposed
\cite{qccluster1,qccluster2,qccluster3}. The distinct advantage of
one-way computing strategy is that it separates the process of
generating entanglement and executing the computation. So one can
tolerate failures during the generation process simply by repeating
the process providing the failures are heralded. On the other hand,
due to its novel application in quantum computing, generation of the
cluster states
\cite{generate1,generate2,generate3,generate4,generate5,generate6}
also attracts many attentions. Here, as a direct application of the
proposed 2-qubit QCPG operation, we also present a way of generating
the multipartite cluster states via SQUIDs trapped in cavity.

\section{Basic models}
The  SQUID considered in this paper is the radio frequency SQUID
(rf-SQUID), which consisting of a Josephson tunnel junction in a
superconducting loop, the size of which is on the order of
10$\sim$100$\mu$m. The rf-SQUID considered here has three-level
$\Lambda$-type energy structure, which includes two lower flux
states $|0\rangle$, $|1\rangle$ and an upper state $|e\rangle$. The
two lower flux states $|0\rangle$ and $|1\rangle$ reside in two
distinct potential valley and serve as logic 0 and 1 in our
proposal. Suppose the coupling of $|0\rangle$, $|1\rangle$ and
$|e\rangle$ with other levels of the rf-SQUID via the cavity is
negligible, which can be readily satisfied by adjusting the level
spacings of the rf-SQUID. For a rf-SQUID, the level spacings can be
easily changed  by varying the external flux $\Phi _x$ or the
critical current $I_c$ \cite{levelspace}. Hence, coupling between
microwave pulse and any particular rf-SQUID qubit can be obtained
selectively, via frequency matching.

\subsection{SQUID driven by classical field}
We firstly review the way of implementing single-qubit operations,
which can be realized by a rf pulse on the rf-SQUID. Let's firstly
consider a rf-SQUID driven by a rf pulse. If it is resonant with the
$|0\rangle\leftrightarrow | 1\rangle $ transition but far-off
resonant with the transitions of $|0\rangle\leftrightarrow|e\rangle$
and $|1\rangle \leftrightarrow|e\rangle$ of the rf-SQUID, the
interaction Hamiltonian is \cite{laserh}
\begin{equation}
H_I=i\Omega(|0\rangle\langle 1|-|1\rangle\langle 0|),
\end{equation}
where $\Omega$ is the Rabi frequency between the levels $|0\rangle$
and $|1\rangle$. Thus a rf pulse with duration time $t$ results in
the following rotation
\begin{eqnarray}
|0\rangle\rightarrow\cos\Omega t|0\rangle -\sin\Omega
t|1\rangle,\nonumber\\
|1\rangle\rightarrow\cos\Omega t|1\rangle+\sin\Omega t|0\rangle.
\end{eqnarray}
Similarly, single-qubit operations in the basis \{$|0\rangle$,
$|e\rangle$\} and \{$|1\rangle$, $|e\rangle$\} can also be realized.
In the rest part of this letter, we will use some single qubit
operations without specifying the interaction time.

\subsection{Cavity-SQUIDs resonant interaction}
We then consider 2 rf-SQUID qubits simultaneously interacting with a
single-mode cavity. The distance of the two qubits is large enough
so that the interaction between the two SQUIDs is completely
negligible. If the cavity mode is resonant with the
$|0\rangle\leftrightarrow | 1\rangle $ transition but far-off
resonant with the transitions of $|0\rangle\leftrightarrow|e\rangle$
and $|1\rangle \leftrightarrow|e\rangle$ of the 2 rf-SQUIDs, the
interaction Hamiltonian can be expressed, in the interaction picture
\cite{laserh}, as
\begin{equation}
\label{1}
H=\Omega_{1}(a^{+}|0\rangle_{1}\langle1|+a|1\rangle_{1}\langle0|)
+\Omega_{2}(a^{+}|0\rangle_{1}\langle1|+a|1\rangle_{1}\langle0|),
\end{equation}
where $\Omega_{1}$ and $\Omega_{2}$ are the coupling strength of the
SQUIDs 1 and 2 with the cavity, respectively. $a^{+}$ and $a$ are
the creation and annihilation operators for the cavity mode. Then we
can obtain the following evolutions
\begin{eqnarray}
|0\rangle_{1}|0\rangle_{2}|0\rangle&&\rightarrow|0\rangle_{1}|0\rangle_{2}|0\rangle,\nonumber\\
|0\rangle_{1}|e\rangle_{2}|0\rangle&&\rightarrow|0\rangle_{1}|e\rangle_{2}|0\rangle,\nonumber\\
|1\rangle_{1}|0\rangle_{2}|0\rangle&&\rightarrow\frac{\Omega_{1}}{\Omega}[\frac{1}{\Omega}(\Omega_{1}\cos
\Omega t+\frac{\Omega_{2}^{2}}{\Omega_{1}})
|1\rangle_{1}|0\rangle_{2}|0\rangle\nonumber\\
&&+\frac{\Omega_{2}}{\Omega}(\cos \Omega
t-1)|0\rangle_{1}|1\rangle_{2}|0\rangle\nonumber\\
&&-i \sin\Omega
t|0\rangle_{1}|0\rangle_{2}|1\rangle],\nonumber\\
|1\rangle_{1}|e\rangle_{2}|0\rangle&&\rightarrow(\cos\Omega_{1}t|1\rangle_{1}|0\rangle-i
\sin\Omega_{1}t|0\rangle_{1}|1\rangle)|e\rangle_{2},
\end{eqnarray}
where $\Omega=\sqrt{\Omega_{1}^{2}+\Omega_{2}^{2}}$.

\section{QCPG operation}
Now, we consider the implementation of the 2-qubit QCPG.

Step 1. Let qubit 2 interact with a classical fields tuned to the
transition $|1\rangle\leftrightarrow|e\rangle$. Choosing the
amplitude and phase of the classical field appropriately, one
obtains the transition $ |1\rangle_{2}\longrightarrow|e\rangle_{2}$.

Step 2. Send the two SQUID qubits simultaneously to the vacuum
cavity.  If we choose $\Omega_1 t=\pi$ and
$\Omega_{2}=\sqrt{3}\Omega_{1}$, which can be achieved by choosing
coupling strengths and interaction time appropriately. Then we have
\begin{eqnarray}
|0\rangle_{1}|0\rangle_{2}\rightarrow|0\rangle_{1}|0\rangle_{2},
|0\rangle_{1}|e\rangle_{2}\rightarrow|0\rangle_{1}|e\rangle_{2},\nonumber\\
|1\rangle_{1}|0\rangle_{2}\rightarrow|1\rangle_{1}|0\rangle_{2},
|1\rangle_{1}|e\rangle_{2}\rightarrow-|1\rangle_{1}|e\rangle_{2},
\end{eqnarray}
where we have omitted the cavity state, which was left in the vacuum
state.

Step 3. Let qubit 2 interact again with a classical field tuned to
the transition $|1\rangle\leftrightarrow|e\rangle$ and one obtains
the transition $ |e\rangle_{2}\longrightarrow|1\rangle_{2}$.

The states of the system after each of the above steps can be
described as
\begin{eqnarray}
|0\rangle_{1}|0\rangle_{2}\longrightarrow|0\rangle_{1}|0\rangle_{2}
\longrightarrow|0\rangle_{1}|0\rangle_{2}\longrightarrow|0\rangle_{1}|0\rangle_{2},\nonumber\\
|0\rangle_{1}|1\rangle_{2}\longrightarrow|0\rangle_{1}|e\rangle_{2}
\longrightarrow|0\rangle_{1}|e\rangle_{2}\longrightarrow|0\rangle_{1}|1\rangle_{2},\nonumber\\
|1\rangle_{1}|0\rangle_{2}\longrightarrow|1\rangle_{1}|0\rangle_{2}
\longrightarrow|1\rangle_{1}|0\rangle_{2}\longrightarrow|1\rangle_{1}|0\rangle_{2},\nonumber\\
|1\rangle_{1}|1\rangle_{2}\longrightarrow|1\rangle_{1}|e\rangle_{2}
\longrightarrow-|1\rangle_{1}|e\rangle_{2}\longrightarrow-|1\rangle_{1}|1\rangle_{2}.
\end{eqnarray}
The above transformations correspond to 2-qubit QCPG operation,
where qubit 1 and 2 are the control and target qubit, respectively.
In this way, a scheme for implementing the 2-qubit QCPG based on
cavity-SQUID system is proposed. Our scheme is achieved without any
type of measurement, does not use the cavity mode as the data bus
and only requires a single resonant interaction of the SQUID-cavity
system.  Thus the presented scheme is very simple and the required
interaction time is very short. The simplification of the process
and the reduction of the operation time are important for
suppressing decoherence.

\section{Generation of cluster states}
Now, we consider the generation of the cluster states. Firstly, we
focus on the 2-qubit cluster state case. Initially, rf-SQUIDs 1 and
2 have been prepared in the state $1/\sqrt{2}(|0\rangle+|1\rangle)$
and the cavity in vacuum state $|0\rangle_C$. Thus the state of the
quantum system is
\begin{equation}
\frac{1}{2}(|0\rangle_1+|1\rangle_1)(|0\rangle_2+|1\rangle_2)|0\rangle_C.
\end{equation}
Repeat the process of the above 2-qubit QCPG operation on rf-SQUIDs
1 and 2 with the aid of the cavity. After the gate operation the
state of the quantum system will evolve to
\begin{equation}
\label{b2}
\frac{1}{2}(|0\rangle_1+|1\rangle_1\sigma_1)(|0\rangle_2+|1\rangle_2),
\end{equation}
where $\sigma_1=|0\rangle_2\langle0|-|1\rangle_2\langle1|$ and we
have omitted the state of the cavity, which is left in the vacuum
state and disentangled with the prepared entangled state. The state
in Eq. (\ref{b2}) is a bipartite cluster state.

Next, we consider the generation of \emph{N}-qubit cluster state.
Assume all the \emph{N} rf-SQUIDs  have been prepared in the state
$1/\sqrt{2}(|0\rangle+|1\rangle)$ and a cavity in the vacuum state.
the initial state of the system is
\begin{equation}
\label{ni}
\frac{1}{\sqrt{2^{N}}}\bigotimes_{i=1}^{N}(|0\rangle_i+|1\rangle_i)|0\rangle_C.
\end{equation}
For rf-SQUID $i$ ($i<N$), repeat the above procedures for generating
bipartite cluster state on the rf-SQUIDs $i$, ($i+1$) and the cavity
(these selective interactions can be embodied via frequency
matching) and end the process when $i=N$. Then the \emph{N}
rf-SQUIDs will be prepared in
\begin{equation}
\label{n}
\frac{1}{\sqrt{2^{N}}}\bigotimes_{i=1}^{(N-1)}(|0\rangle_i+|1\rangle_i\sigma_i)(|0\rangle_N+|1\rangle_N),
\end{equation}
with
$\sigma_i=|0\rangle_{(i+1)}\langle0|-|1\rangle_{(i+1)}\langle1|$.
The state in (\ref{n}) is the \emph{N}-qubit cluster state and we
have omitted the state of the cavity, which is disentangled with the
cluster state.

For the case of the inevitable imperfectness of the experimental
exploration of our scheme, it is very reasonable that one can only
generate the cluster state of a certain length we label this
critical number as $n_c$. To generate a cluster chain of a length
$n>n_c$, we can simply parallelling generate cluster chains of
length under the critical number and then fuse them together to
further increase its length. This idea can also be perfected even if
the QDPG is not deterministically \cite{generate4}. In Ref.
\cite{generate4}, Duan et al. also generalized the idea to the
generation of two-dimensional square lattice cluster states from a
set of cluster chains with QCPG only succeed with an arbitrarily
small probability. The two-dimensional square lattice cluster state
prepared at numerous qubits is a universal "substrate" for quantum
computation. After the preparation of the states, the remaining work
is only to perform single-qubit measurements, and the final results
are read out from those qubits that were not measured in the whole
process. In this sprit, many recent schemes
\cite{generate4,generate6,c} have shown how to use probabilistic
gate operations  to construct entangled states with certainty. Our
scheme is of deterministic nature, thus it holds more promising
future.

\section{Discussions and conclusion}
Before ending the paper, we briefly address the experimental
feasibility of the proposed scheme. Among the variety of systems
being explored for hardware implementations for quantum computation,
the cavity-SQUID system, as well as other solid state circuits, is
favored because it is easy embedded in electronic circuits and
scaled up to large registers \cite{coherent}, and the control and
measurement techniques are quite advanced \cite{rsquid}. The
interaction time can be perfectly controlled by external control.
The time constants involved are long enough to realize all the
required manipulations \cite{flux}. Finally, coupling between
microwave pulses and any particular rf-SQUID qubit can be obtained
selectively via frequency matching. Thus our scheme might be
realizable within current technology.

For the sake of definitiveness, let us estimate the experimental
feasibility of realizing the logic gates using SQUIDs with the
parameters already available in present experiment
\cite{experiment1,experiment2,experiment3}. Suppose the quality
factor of the superconducting cavity is $Q=1\times10^{6}$ and the
cavity mode frequency is $\omega_c=50$GHz, the cavity decay time is
$k^{-1}=Q/\omega_c=20$$\mu$s. The realistic length of the cavity is
$l=1.4$mm, when the two SQUIDa are located each at one of the
antinodes of the cavity, the distance of them, $D$, is equal to the
length of the cavity. For SQUIDs of size $d=40$$\mu$s, we get
$D/d=35$ which perfectly satisfies the requirement of $D\gg d$. The
upper state energy relaxation constant is $\gamma_e^{-1}=2.5$$\mu$s.
For a superconducting standing-wave cavity and a SQUID located at
one of antinodes of the magnetism field, the coupling constant is
$g=1.8\times10^{8}$Hz, thus the resonant interaction time for our
scheme is $T_r=\pi/g\simeq1.7\times10^{-8}s$. Meanwhile, the strong
coupling condition can be perfectly achieved as $g^{2}/\gamma_e
k=1.6\times10^{6}\gg1$. The Rabi frequency for the interaction of
the rf pulse and the SQUID is $\Omega=8.5\times10^{7}$Hz, thus the
interaction time for the single photon detection is
$T_l=\pi/2\Omega\simeq1.8\times10^{-8}s$. Both interaction time
($T_r$ and $T_l$) are much shorter than the cavity decay time and
the relaxation time of the upper state. In addition, the above
estimation is very conservative compare to present experimental
achievements, in fact, a superconducting cavity with
$Q=3\times10^{8}$ has already been reported \cite{hq}.

In conclusion, we have investigated a simple scheme for implementing
the 2-qubit QCPG gate based on cavity-SQUID systems. The presented
schemes are achieved without any type of measurement. Our scheme is
also a deterministic one without any auxiliary SQUID serving as
data-bus. In addition, the implementation of our scheme is simple,
which is very important in view of decoherence and the successful
probability and the fidelity both reach unit. We also roughly
estimated the experimental feasibility of our schemes, which shows
they are well within current techniques, thus our suggestion may
offer a simple way of implementing the quantum logic gate via
QED-SQUID system. As an direct application of the QCPG operation, we
also proposed a scheme for generating the cluster states, which are
universal "substrate" for quantum computation.

\acknowledgements This work is supported by the Key Program of the
Education Department of Anhui Province (No. 2006kj070A), the Talent
Foundation of Anhui University and for Z.-Y. Xue by the Postgraduate
Innovation Research Plan from Anhui University.

\end{document}